\documentclass[manuscript,screen]{acmart}
\AtBeginDocument{%
  }

\setcopyright{acmlicensed}
\copyrightyear{2018}
\acmYear{2018}
\acmDOI{XXXXXXX.XXXXXXX}
\acmISBN{978-1-4503-XXXX-X/2018/06}

\newcommand{\textcite}[1]{\citeauthor{#1} \cite{#1}}

\usepackage{tabularray}
\usepackage{xcolor} %to color text red for the final draft

\newif\ifanonymous
\anonymousfalse

\begin{document}

\title{Piloting Planetarium Visualizations with LLMs during Live Events in Science Centers}

% A few title options:
% Interaction Copilots: Using conversational agents for steering visualization during live events in science centers
% Challenges and Opportunies of [LLM copilots/autonomous agents] in visualization during live events
% Elliciting science center guides' use of LLM agents for interaction during live shows.


\ifanonymous
\else

\author{Mathis Brossier}
\email{mathis.brossier@liu.se}
\orcid{0000-0001-7653-9457}
\affiliation{%
  \institution{Linköping University}
  \country{Sweden}
}

\author{Mujtaba Fadhil Jawad}
\email{mujtaba.fadhil.jawad@liu.se}
\orcid{0009-0001-5939-5495}
\affiliation{%
  \institution{Linköping University}
  \country{Sweden}
}

\author{Emma Broman}
\email{emma.broman@liu.se}
\orcid{0009-0006-0072-5134}
\affiliation{%
  \institution{Linköping University}
  \country{Sweden}
}

\author{Ylva Selling}
\email{ylva.selling@liu.se}
\orcid{}
\affiliation{%
  \institution{Linköping University}
  \country{Sweden}
}

\author{Julia Hallsten}
\email{julia.hallsten@visualiseringscenter.se}
\orcid{https://orcid.org/0009-0007-0704-8823}
\affiliation{%
  \institution{Visualisering Center C}
  \country{Sweden}
}

\author{Alexander Bock}
\email{alexander.bock@liu.se}
\orcid{https://orcid.org/0000-0002-2849-6146}
\affiliation{%
  \institution{Linköping University}
  \country{Sweden}
}

\author{Johanna Björklund}
\email{johanna@cs.umu.se}
\orcid{0000-0003-0596-627X}
\affiliation{%
  \institution{Umeå University}
  \country{Sweden}
}

\author{Tobias Isenberg}
\email{tobias.isenberg@inria.fr}
\orcid{0000-0001-7953-8644}
\affiliation{%
  \institution{Inria}
  \country{France}
}

\author{Anders Ynnerman}
\email{anders.ynnerman@liu.se}
\orcid{0000-0002-9466-9826}
\affiliation{%
  \institution{Linköping University}
  \country{Sweden}
}

\author{Mario Romero}
\email{mario.romero@liu.se}
\orcid{0000-0003-4616-189X}
\affiliation{%
  \institution{Linköping University}
  \country{Sweden}
}

\author{Lonni Besançon}
\email{lonni.besancon@gmail.com}
\orcid{0000-0002-7207-1276}
\affiliation{%
  \institution{Linköping University}
  \country{Sweden}
}

\fi

\renewcommand{\shortauthors}{Brossier et al.}

\begin{abstract}

We designed and evaluated an AI pilot in a planetarium visualization software, OpenSpace, for public shows in science centers. The piloting role is usually given to a human working in close collaboration with the guide on stage. 
We recruited 7 professional guides with extensive experience in giving shows to the public to study the impact of the AI-piloting on the overall experience.
The AI-pilot is a conversational AI-agent listening to the guide and interpreting the verbal statements as commands to execute camera motions, change simulation time, or toggle visual assets. 
Our results show that, while AI pilots lack several critical skills for live shows, they could become useful as co-pilots to reduce workload of human pilots and allow multitasking. We propose research directions toward implementing visualization pilots and co-pilots in live settings.

\end{abstract}



%\ccsdesc[500]{Do Not Use This Code~Generate the Correct Terms for Your Paper}
%\ccsdesc[300]{Do Not Use This Code~Generate the Correct Terms for Your Paper}
%\ccsdesc{Do Not Use This Code~Generate the Correct Terms for Your Paper}
%\ccsdesc[100]{Do Not Use This Code~Generate the Correct Terms for Your Paper}



%\keywords{Do, Not, Use, This, Code, Put, the, Correct, Terms, for, Your, Paper}

\received{20 February 2007}
\received[revised]{12 March 2009}
\received[accepted]{5 June 2009}

\maketitle

\section{Introduction}

The development of LLMs and foundation models capable of reasoning about complex tasks facilitates a natural interaction between machines and humans. Uses of AI assistants are flourishing in various applications of visualization and interaction \cite{brossier2026star}, such as data science \cite{chen_interchat_2025}, operative surgery \cite{hao_enhancing_2025}, forensics sciences \cite{Pooryousef_autopsy_AI}, or science communication \cite{jia_voice_2025}. LLM applications are also becoming ubiquitous in academia \cite{sajjadi_mohammadabadi_survey_2025} and in our broader society \cite{dreksler_what_2025,chkirbene_large_2024}. This development, however, remains controversial within the scientific community \cite{binz_how_2025} and the public at large \cite{eom_societal_2024}. Beyond social and societal considerations, the interplay between users and LLM assistants is not well understood, especially when they exhibit anthropomorphic features \cite{schlesener_am_2025}.
A key limitation of current AI-naturalness in collaboration with humans is the inability of LLMs to be proactive in a conversation \cite{chatterjee_proactllm_2025}: an LLM only generates a response after a user query, leading to a dialogue asymmetry. Stock LLM assistants sometimes confidently state that ``they will process the user request in the background, and notify them when it is done.'' Classical LLM assistants, however, do not have such autonomy and can only generate text and reasoning when prompted. Research is ongoing to improve the \emph{reasoning} autonomy of LLMs \cite{ferrag_llm_2025}, but more research is needed to explore \emph{interaction} autonomy.

During educational and interactive live shows, a human guide and visualization pilot collaborate to deliver an interactive and immersive experience to a large audience. 
In this study, we build upon the work of \textcite{tabalba_articulatepro_2025} and \textcite{jia_voice_2025} who studied how proactive LLM agents can support human guides interventions in science centers during live events. Working toward this goal, we implemented LLM-based interaction with OpenSpace, an astrophysics visualization software commonly used in live shows involving a presenter and a visualization pilot\cite{bock_openspace_2019}. To further explore the topic of pragmatics and proactive AI, we compare two test cases: a \emph{reactive} mode, in which the AI agent only responds to explicit and directed queries; and a \emph{proactive} mode, in which the agent intervenes when it deems necessary to perform implicit queries, while the human guide is talking. We conducted a comparative study between the two modes with five professional planetarium guides and we interviewed two additional experts to understand human visualization piloting and presentation during live events. Noteworthy, the experts have extensive experience with human pilots and are able to provide comparative observations against the two AI conditions. We explore the strengths and weaknesses of both AI modes and contribute research directions stemming from the study and participants' experiences as both public guides and visualization pilots.
We center our exploration around three research questions.
\begin{itemize}
    \item RQ1: What are the strengths and weaknesses of AI co-pilots compared to human co-pilots in live show settings?
    \item RQ2: What are the strengths and weaknesses of proactive AI pilots compared to reactive AI pilots?
    \item RQ3: What further developments and research directions should be explored for AI co-pilots in visualization?
\end{itemize}

\section{Related work}

%\paragraph{Conversational AI in visualization.}

% Lonni: You also need to cite your poster here and make sure that you show that the implementation work is not the focus.
% Like Voice \cite{jia_voice_2025}, we can do a double analysis: dichotomous interpretations for explicit queries, and qualitative interpretations for implicit queries.

Different dimensions of using conversational agents in visualization have been studied \cite{brossier2026star}. Most prominently, using natural language to query data, (e.g., text2sql tasks \cite{hong_next-generation_2025}) and visualization recommendation (e.g., generation of visualization via code \cite{basole_generative_2024}) has been studied extensively, including work focusing on LLM approaches \cite{hong_next-generation_2025}. Visualization reading has also been studied \cite{davila_chart_2021}. Much rarer is the work focusing on natural language navigation using LLMs \cite{jia_voice_2025}. 

Applications of proactive (or autonomous) agents are emerging. While the terminology is not yet fully established, we chose the term ``proactive,'' as it counter-balances well with ``reactive.'' The term is used in context-aware models \cite{yang_contextagent_2025}, video streaming models \cite{zhang_eyes_2025}, as well as in applications to visualization \cite{tabalba_articulatepro_2025}. The first workshop focusing on the topic of proactive LLMs occurred in 2025 \cite{chatterjee_proactllm_2025}. We take inspiration from the work by \textcite{tabalba_articulatepro_2025}, who studied proactive agents in visualization for collaborative work, and \textcite{jia_voice_2025}, who studied visualization navigation with conversational agents. We combine these approaches, and therefore study proactive visualization navigation during live educational shows about the solar system.

% ArticulatePro \cite{tabalba_articulatepro_2025} has a proactive vs. non-proactive LLM assistant with interesting findings, such as: The user preference goes to the case tried first, 

\section{System implementation}
Our system is inspired by previous LLM-steered visualization implementations by Brossier et al. \cite{brossier_space_2024}. We built an LLM-supported system with two operation modes. In \emph{reactive} mode, the guide holds a manually triggered microphone and can utter voice commands to control the visualization software. In \emph{proactive} mode, the microphone is constantly capturing the guide's voice while the LLM interprets and executes visualization commands proactively.
We base our work on the astrophysics visualization software OpenSpace \cite{bock_openspace_2019}, used notably in planetariums during ``tour of the universe'' live events with broad audiences but which has generic rendering and explaining capabilities including, e.g., molecular dynamics in contexts \cite{BrossierMoliverse}. We control OpenSpace via its Lua API over WebSocket. We use a separate JavaScript process to coordinate between an LLM agent, OpenSpace's API, and peripherals (microphone, remote trigger and user interface). We employ OpenAI's low-latency, multimodal model (\texttt{gpt-realtime-2025-08-28}) for the natural language reasoning tasks. The LLM model is capable of dispatching \emph{tool calls} to travel to different scene nodes, toggle the visibility of nodes, travel to a geolocation (lat, long, altitude), and change the simulation time and speed. While our first implementation iteration performed well at the start of conversations, it typically lost track of the instructions after a few turns. To mitigate this issue, we added few-shot examples as warm-up at the start of conversations. We deleted model-generated textual messages and added a short system instruction reminder at the tail of the conversation thread.

In \emph{reactive} mode, we feed the audio input from the microphone directly to the LLM. In \emph{proactive} mode, however, we employ a \emph{streaming} speech-to-text model developed by \citeauthor{machacek_simultaneous_2025} \cite{machacek_simultaneous_2025}, which allows us to prompt the low-latency LLM continuously with the guide spoken utterances translated to text. In both modes, the LLM does not produce any textual or audio output in addition to dispatching tool calls. This setup imitates the setup between the human pilot and guide, where typically only one-way verbal communication from the guide to the pilot is possible. The pilot replies through interpreting the commands which become view transformations in the visualization. When the LLM judges that no action needs to be performed, it replies with a call to a special ``no-op'' tool call which we ignore.

\section{Study methodology}

% \ti{[somehow the story part I am missing is the whole setup: the technical implementation, or the use of an LLM-based system created by someone else, then the experiences with it, plus some input/intuition that different forms of engagement by the LLM (\emph{reactive} vs. \emph{proactive}) may lead to different user experiences, etc.; from that then you need to extract the need for a study, and the research questions you want to ask or the hypotheses you want to test; only then can you discuss the best methodology and study protocol for the experiment]}

% \ti{[don't you need to say something here not only about the location but also about the technical setup, i.e., how the LLM is connected to which visualization system? at least on a high level, before you explain things more in detail below.]}
Our study consisted of five one-hour sessions, one for each of five participants in a full-dome planetarium at \ifanonymous <anonymized location> \else  the Visualization Center C in Norrköping, Sweden\fi. We compared usage of our system of two cases, (\emph{reactive} vs. \emph{proactive}) against the human baseline, which we established through interviews with two additional consulting experts. We collected both quantitative data from the system usage and qualitative data from participant feedback. We performed a thematic analysis from the interviews and an error-case analysis from the system operation logs (\autoref{sec:error-case}).

% \textcolor{orange}{I'm not sure if I should invest in, or will have time to do that rigorously? And I have some missing data for the first 3 participants, and Emma was inconclusive, so basically I only have robust data for Ylva. This aspect definitely needs more work, but maybe this is not the paper for it}.

\subsection{Participants}

We recruited five expert participants who are professionals working at a science center. All the participants had experience presenting in the dome to public audiences. Two (P1 and P2) were educators at the facilities and had the most experience engaging with visitors, but a limited technical background. The other three (P3--P5) were engineers and researchers working on the dome software development and had also presented several dome shows. In addition, we recruited two consulting experts with leading roles in science communication (C1) and dome software development (C2). They provided us with qualitative feedback during a longer interview on a typical dome show experience with a public audience, including elements that make shows engaging, guide $\leftrightarrow$ pilot teaming, planning of shows, and their opinions on AI. %(\autoref{sec:baseline}). %this section is currently commented out.
All participants are working at our institution and are exempted from requiring ethical approval or getting paid to participate in the study. As is commonly the case \cite{pooryousef2025lessons} in studies involving domain experts contributing their time (e.g., \cite{besancon:hal-01795744,besancon:hal-02381513,Meyer,pooryousef_criminator_2026,Rogers,Sedlmair}), co-authorship was offered to the participants after their individual participation.
% \ti{[need ethics statement, need informed consent, need payment status]}

\subsection{Study protocol}

We ran one-hour sessions with each participant (P1--P5) in the dome. We asked the participants to conduct two simulated dome shows to be able to compare the two interaction modalities with the LLM assistant. %(\emph{reactive} vs. \emph{proactive}).
We followed with a 20-minute semi-structured interview. Due to some instability with the software, the first author had to intervene several times during the presentation. For participant P4, we could not reliably run the software, so we instead decided to have them be spectator of P5's experience and we interviewed both participants together.

The simulated show was a shortened version of a typical show, which consists of a ``tour of the Solar System'', starting from a location on Earth and visiting the moons and planets of the Solar System. We gave participants freedom to explore and discuss to let them understand and adapt to the system's functionality and limitations. We compared two system uses: in \emph{reactive} mode, the guides held a hand-triggered microphone which they could use to perform direct voice commands, such as ``Go to the Moon.'' In \emph{proactive} mode, the system was constantly reacting to the guide's speech, without an explicit trigger. For example, mentioning the Moon would initiate a flight to it. Each test case lasted up to 10 minutes. We alternated the starting case between participants to limit bias. P1, P3, and P5 started with the \emph{reactive} case, P2 and P4  with the \emph{proactive} case. After the first test case, we conducted a short ($\approx$ 5 minutes) unstructured interview to collect the participant's immediate feedback while preparing the system for the next test case. After the second case, we conducted a 20-minute semi-structured interview.

\subsection{Data collection and analysis}

We collected both data from the system logs, as well as the qualitative feedback from the participant interviews. From the system we logged the session audio and transcripts, the end-to-end latency, and the nature, number and success rate of system interventions. At this stage, we focus on  reporting on the insights from the interviews.

\section{Results}

\subsection{Comparison with a human pilot}

Regarding their expectations and enjoyment, some participants expressed clear enjoyment. It is a ``cool concept'' (P1), ``more fun than anticipated. [The system is] listening to me and [doing] stuff as I am talking'' (P2). We then asked participants about the current and projected capabilities of AI pilots in dome shows. All participants agree that AI pilots are missing many nuances compared to humans, which explains that while they sometimes behave correctly, they are far from having the robustness to sustain a full show autonomously. P1: ``A human pilot can guess what I want to say and anticipate''. P2 adds: ``sometimes we open questions from the audience [\dots]. With specific things it may fail''.

%We were intrigued about what makes the piloting role fundamentally human. All participants, especially guide experts (P1, P2, C1), emphasized that acknowledging the pilot during the show is essential. 
We examined what makes the piloting role fundamentally human. Expert participants (P1, P2, C1), emphasized the importance of explicitly acknowledging the pilot during the show.
Being a live event from both the guide and the pilot sides, there is ``something to it'' as a show for an audience (P2). It ``underlines that it's a live presentation'' and it ``compensates for the limited interaction with the audience'' (C1). Two participants (P1, P3) also explained that the AI lacked proper \emph{pacing}: ``it's all about the pace. If I talk about the ISS for 5 minutes it's going to be boring. But the first 5 minutes traveling slowly there would make it more interesting. Buildup'' (P3). The pilot is constantly moving the camera, like driving a car. As in movies, the scene composition and motion also supports storytelling and immersion, and camera control must be precise, as jittering or too fast movements can cause discomfort. Due to their poor temporal awareness and their inherent latency, LLM systems struggle with  reactive, fluid, and precise camera control.

We further asked about the relationship between the guide, the pilot and the public. There are only limited interactions possible, but they are nonetheless important. The guide expert (C1) explained that perceiving the audience's engagement is possible, yet hard to formalize. A strong signal is the breathing (gasps, silences), as well as feeling ``something in the air.'' Reaction to jokes is the easiest way to probe for engagement. Such cues are difficult to measure for any machine. From the booth behind the stage where the pilot sits, it is very hard to read anything from the audience.

\subsection{Comparison between reactive and proactive cases}

The preference between the two modes was mixed. Due to intermittent system failures, it was sometimes difficult for participants to imagine the best-case scenario. P1 preferred the proactive mode, as it felt more like interacting with a human because of reading context cues. P2, P3, and P5 preferred the reactive mode because it was more predictable and reliable.
The reactive mode was more reliable, but added to the mental load of participants. There was better flow without direct questions and addressing to the AI. Further, remembering to press the trigger increased cognitive load which made it harder to sustain conversation flow (P1, P2). The proactive mode thus felt more fluid or natural.

% // preference
% (P1) preferred proactive because it felt more like interacting with a human because of reading context cues. Didn't have to do stops.
% (P2) preferred the reactive mode, as it was more reliable. But P2 added that there is a better flow without direct questions, and that addressing to the AI, as well as remembering to press the trigger, increased cognitive load which made it harder to keep the conversation flow going.
% (P3) I would have more control with the switch but I like the idea of it listening to me. [in proactive mode] it's too sensitive in switching between objects.

\subsection{Combining human and AI}

A future approach could lie in piloting cooperation between a human and an AI agent. Three directions emerged from the discussion with the participants. First, we established that AI was rather good at toggling assets based on context, while it performed very poorly with camera and pace control. Conversely, human pilots have to simultaneously steer the camera and toggle assets. An AI co-pilot could help by preparing the next actions ahead of time. To illustrate, P3 shared an anecdote: ``[the public was] excited about satellites which we didn't prepare for. I was frantically adding assets about satellites. It would be great if AI could add all of them at once''.

% (P1) perspective on AI: slightly worried. ai should do the "boring stuff" but it often is doing the "fun stuff" in place of humans. The pilot in dome shows adds intrinsic value, and it would be sad if it was deleted. 

% (P2) sees no clear opportunity for combining a human and an AI.

% // workload
% (P3) when I do earthrise, I want to start a vlc video and at the same time start something in openspace. It could help a lot. We have a lot of workload as copilots. 4 panels open at the same time. I have a manuscript and I start enabling things ahead. 
% (P4) [high workload] when there is a lot of things happening at the same time, when I have to travel precisely. ``sometimes I would like to 2 things do at the same time.''
% (P5) [high workload] when the thing is not in the right space, when I have to change time.

% The hard part for pilots is the fine control of the camera. But enabling/disabling things is easy (P3, P5).

% (P3) anecdote: aviation ppl were excited about satellites. I was frantically adding assets about satellites. It would be great if ai could add all satellites.

\label{sec:error-case}
\subsection{Error case analysis}

Action correctness is a spectrum. One may argue that a direct query, such as ``Show me the Earth''  is objective and there is no interpretation possible. In 3D visualization software, however, the level of zoom and the location in focus have a definitive impact on the quality of the action performed. Indeed, the system would frequently zoom in on a location that lies on the dark side of the planet, or is too zoomed in/out to show an overview or the details of interest. A qualitative analysis of error cases (both their characteristics, and prevalence) is needed.
From our observations, we characterized a spectrum comprising the following error dimensions for AI pilot actions, with failure examples:

\begin{itemize}
    \item \textbf{Detection.} How sensitive and specific action detections are.
    \emph{Failure example: Mishearing the guide.} 
    \emph{Means of evaluation: counting true positives, false positives, and guide intents.}

    \item \textbf{Reasoning.} How well the system understands users' implicit or explicit intent.
    \emph{Failure example: Asking to show the Moon ``as we see it from Earth''.} 
    \emph{Means of evaluation: can use reasoning benchmarks} \cite{lu2025toolsandbox}. 
    
    \item \textbf{Context.} Contextual elements include current visualization state, audience, and conversation history.
    \emph{Failure example: Showing the dark side of a planet} 
    \emph{Means of evaluation: reasonably can only be coded by humans, but can be trained on examples.}

    \item \textbf{Naturalness.} Whether the action feels human-like.
    \emph{Failure example: Poor steering of the view.} 
    \emph{Means of evaluation: subjective human assessment.}
\end{itemize}

\section{Directions for future work}

In light of our experiment and interviews, we establish the following directions for future work.

\paragraph{Visualization interaction recommender systems.} 

A clear feedback from the participants was that AI pilots are nowhere near capable of replacing human pilots (and it was not the purpose of the study). However, they may prove useful as recommender systems to, for example, speed up human pilots and reduce their cognitive load by digging through assets and simplifying the onboarding of novice pilots. Visualization recommender systems (i.e., capable of producing visualizations based on user requests), including natural language and LLM-based ones, have been studied extensively \cite{vartak_towards_2017}. Recommender systems for \emph{interaction} specifically, have, to our knowledge, not been studied before. Such \emph{interaction recommender systems} would consist of an interface which suggests triggerable actions based on the current context, possibly including the guide's speech and, further into the future, gestures and the system state.

\paragraph{Multitasking with multimodality.} 

Participants highlighted that multitasking was frequent during shows. Typically, the pilot would control the camera path and prepare the upcoming assets simultaneously. This problem is amplified by the need for constant motion. In fact, continuous rotation around objects improves depth perception (P4) and engages the audience. To facilitate multitasking, OpenSpace's camera motion has inertia, which allows the pilot to set it in motion and let go of the controls shortly while focusing on another task. When asked about their opinion on multimodal keyboard/mouse and speech interaction, P4 and C2 indicated that the current workflow is sufficient and that speaking out loud in the booth would disturb the audience.

\paragraph{Use of AI during show preparations.} 

Live events are always planned ahead for both the pilot and the guide. According to P3 and C2, live shows are composed of small bricks of storytelling assembled together. Adapting to slightly different shows is easy for the pilot. Designing new assets and storylines, however, requires more work ahead of time and knowledge of the software. An AI pilot could help during the testing and brainstorming in a low-risk environment.

\section{Conclusion}

We explored the concept of LLM visualization pilots for guided interactive presentations. Our study indicates that, while such pilots are not, and likely never will be, replacements for human counterparts, they may find uses as co-pilots, with aims of reducing the cognitive load, enabling multitasking, and facilitating on-boarding of new pilots. Such systems, functioning in the background, need greater levels of autonomy than typical conversational chatbots, for which interaction patterns are too obtrusive in live events. Autonomous LLMs appear as an interesting research direction to facilitate discrete interaction with visualizations by guides during live events.

\begin{acks}
\ifanonymous Acknowledgment and funding information removed for blind reviews. \else 
This work was partially supported by the Marcus and Amalia Wallenberg Foundation (2023.0130), and the Knut and Alice Wallenberg Foundation (2019.0024). We thank NVAB and all the dome-supporting crew at Visualisering Center C. \fi
\end{acks}

\bibliographystyle{ACM-Reference-Format}
\bibliography{references}

\end{document}